\documentclass[prb,twocolumn,longbibliography]{revtex4-1} 

\usepackage{amsmath} 
\usepackage{amssymb} 
\usepackage{wasysym} 
\usepackage{physics} 
\usepackage{graphicx} 
\usepackage{float} 
\usepackage{hhline}
\usepackage{hyperref} 
\usepackage{xcolor} 
\usepackage{lipsum} 
\usepackage{natbib}

\graphicspath{{./figures/}} 
\newif\ifeps 
\epstrue
\epsfalse

\newif\ifbibtex 
\bibtextrue

\begin{document}
\title{Little bits of diamond:\\Optically detected magnetic resonance of nitrogen-vacancy centers}

\author{Haimei Zhang}
\author{Carina Belvin}
	\altaffiliation{Current address: Department of Physics, Massachusetts Institute of Technology, Cambridge, MA 02139, USA}
\author{Wanyi Li}
	\altaffiliation{Current address: Department of Management Science and Engineering, Stanford University, Stanford, CA 94305, USA}
\author{Jennifer Wang}
	\affiliation{Department of Physics, Wellesley College, Wellesley, MA 02481, USA}
\author{Julia Wainwright }
	\affiliation{Department of Physics, Wellesley College, Wellesley, MA 02481, USA}	
\author{Robbie Berg}
	\email[Corresponding author. Email: ]{rberg@wellesley.edu}
	\affiliation{Department of Physics, Wellesley College, Wellesley, MA 02481, USA}

\author{Joshua Bridger}
	\affiliation{Dover Sherborn High School, Dover, MA 02030, USA}

\date{\today}

\begin{abstract}
We give instructions for the construction and operation of a simple apparatus for performing optically detected magnetic resonance measurements on diamond samples containing high concentrations of nitrogen-vacancy (NV) centers. Each NV center has a spin degree of freedom that can be manipulated and monitored by a combination of visible and microwave radiation. We observe Zeeman shifts in the presence of small external magnetic fields and describe a simple method to optically measure magnetic field strengths with a spatial resolution of several microns. The activities described are suitable for use in an advanced undergraduate lab course, powerfully connecting core quantum concepts to cutting edge applications. An even simpler setup, appropriate for use in more introductory settings, is also presented.
\end{abstract}

\maketitle 

\section{Introduction}
\begin{figure}[!htbp]
\includegraphics[width=2.0in]{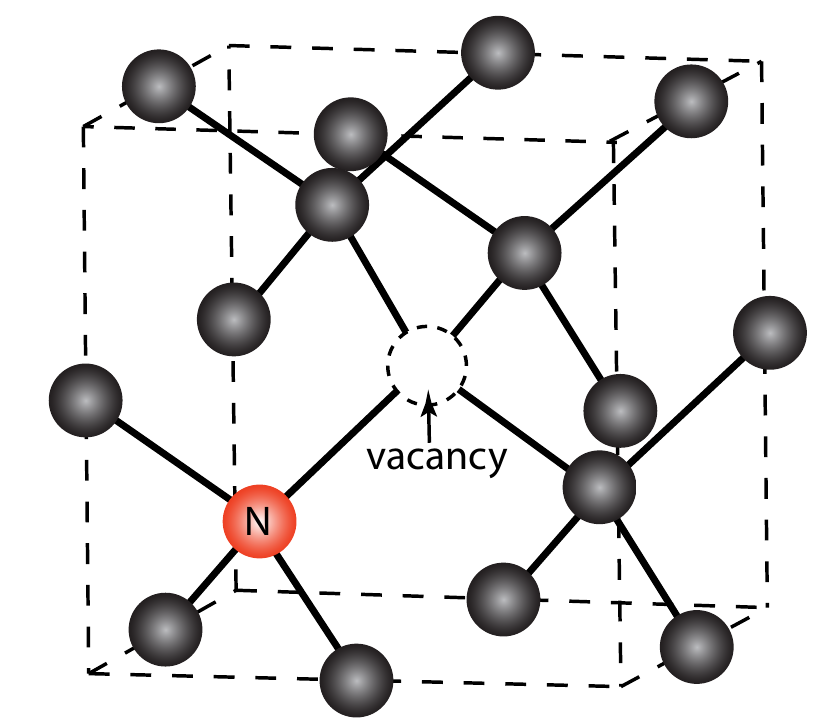}
\caption{An NV center, consisting of a nitrogen atom, which substitutes for a carbon atom, and an adjacent lattice vacancy. }
\label{fig:NV}
\end{figure}

Magnetic resonance spectroscopy is a technique of great power and scope. As a result, magnetic resonance experiments are a common and valuable addition to an advanced undergraduate physics laboratory.\cite{donnally1963some, biscegli1982advanced} Here we describe a series of magnetic resonance experiments where we transfer the detection of the magnetic interactions to the optical domain, greatly increasing  the detection efficiency. The increased sensitivity of this optically detected magnetic resonance (ODMR) markedly simplifies the experiments, allowing students to build much of the setup from scratch. As an added benefit, the system that we study, the nitrogen vacancy (NV) point defect in diamond, has attracted wide attention for its potential application in quantum computing and magnetic sensing applications. As shown in Fig.~\ref{fig:NV}, an NV center consists of one substitutional nitrogen defect and an adjacent vacancy. The electrons that comprise this system are contributed by the nitrogen impurity and the dangling bonds from the carbon atoms that surround the vacancy. Collectively, these electrons possess a net spin of one unit of angular momentum, which can be manipulated and monitored in an ODMR experiment.

To appreciate the power of ODMR, consider this: In a conventional magnetic resonance experiment the signal generated by a single spin is ridiculously small, so, of necessity, experiments are performed on large \textit{ensembles} of spins. For example, a modern NMR spectrometer typically has a minimum sample volume of $\sim 10 \;\mu$L, corresponding to $\sim10^{17}$ (in this case, nuclear) spins. \ifbibtex{\footnote{This is the specification for the Bruker 500 MHz NMR spectrometer that is the workhouse used in organic chemistry labs at Wellesley College.}\else \cite{footnote1} \fi It is therefore quite remarkable that, in recent years, ODMR experiments routinely control and detect the electronic spin state of \textit{individual} NV centers.\cite{Gruber-Science1997, hanson2007spins, childress2007coherent, ChildressPhysicsToday2014} This 17 orders of magnitude improvement in sensitivity is not only amazing, but also of great practical interest, since single spins can serve as quantum bits (qubits) for quantum computing applications\cite{ladd2010quantum} or as atom-sized classical bits for memory storage.\cite{doi2014deterministic} They can also be used to measure a number of physical quantities, such as magnetic field\cite{maze2008nanoscale, balasubramanian2008nanoscale} and temperature,\cite{schirhagl2013nitrogen} with nanometer spatial resolution.

While ODMR experiments that manipulate the spin of a single NV center have become commonplace in research labs around the world, these experiments are challenging, since they require isolating the fluorescence from a single center. This is typically accomplished with a custom confocal microscope, which is expensive, complicated, and requires highly precise alignment.\cite{patange2013instrument} Such an instrument is clearly out of scale for our pedagogical purposes here. Instead we describe ODMR measurements that are made on large ensembles of NV centers. By relaxing the need to isolate a single center, the experiments become \textit{much} easier, making them accessible to students in an undergraduate lab, while still serving as a good introduction to the central ideas in this field.
 
Research on NV centers in diamond is situated in a broad effort to develop new materials and approaches for next generation electronic devices. The current generation of electronics relies largely on silicon-based circuits, but as the size of the features in these circuits decreases, this technology is fast approaching fundamental limits.\cite{markov2014limits} As a result, there is an extensive effort afoot to identify new materials and new approaches that will allow electronics to pass beyond the limits of silicon and other conventional semiconductors. One promising route that has emerged in recent years makes use of ``quantum materials,'' which tend to leverage, rather than battle, the effects of making things small. Diamond NV centers are one prominent example of a quantum material.  \ifbibtex{\footnote{Other examples of quantum materials include single atomic layer materials such as graphene,\cite{geim2013van} which could enable ultrafast signal processing, and topological insulators,\cite{moore2010birth} which may lead to error-free channels for transporting spin information. Developing next generation electronic devices based on quantum materials is the mission of the National Science Foundation supported \textit{Center for Integrated Quantum Materials} (CIQM), which is supporting the work described here. See \texttt{ciqm.harvard.edu} for more information.}\else \cite{footnote2}\fi In this paper we give instructions for the construction and operation of a custom-built fluorescence microscope that can be used to make optically detected magnetic resonance measurements on diamond samples containing high concentrations of nitrogen-vacancy (NV) centers. The activities described here are suitable for use in an advanced undergraduate lab course, effectively connecting core quantum concepts to cutting edge applications. In designing these activities we have been guided by our belief that participating in the construction of the instruments used in our teaching labs is a powerful pedagogical tool.\cite{resnick2000beyond} 

\section{Nitrogen vacancy centers in diamond }

\subsection{Charge state}
An NV center can be found in two different charge states: a negative charge state (NV$^-$) and a neutral state (NV$^0$). The charge state of the defect is imposed by surrounding impurities and the position of the Fermi level. Here we focus on the NV$^-$ defect, since it is the state that exhibits a strong ODMR signal and since it is dominant in the samples we will use. We will refer to it simply as the NV center. The extra negative charge adds to the five electrons associated with the three dangling carbon bonds and two valence electrons from the nitrogen atom, so that there are six electrons associated with the in NV$^-$ center.

\subsection{Optical properties}

A perfect diamond crystal would be colorless and transparent to visible light, since its $5.5\; \rm{eV}$ band gap is greater than the energy of a visible photon. But point defects in a diamond's crystal structure can lead to ``color centers'' -- atom-sized regions that can absorb and emit visible light. Color centers in diamond have a long and rich history, much of it connected to the commercial value of diamond. For example, the famous \textit{Hope Diamond} takes on its beautiful blue coloring because of trace amounts of boron within its crystal structure. Single boron atoms can substitute for carbon atoms in the diamond lattice, forming a color center that can absorb light from the red portion of the spectrum. When white light shines on this diamond, the transmitted light is blueish. Similarly, an NV center creates a color center in diamond. When a sample containing NV centers is illuminated with green light a red fluorescence is observed. As a result, diamonds with high concentrations of NV centers take on a lovely pinkish hue and can be quite valuable as gemstones. But our present interest in NV centers arises not from the aesthetics of these optical properties but from far more practical concerns.

Fortuitously, the spin associated with an NV center interacts strongly with both visible light and microwave radiation, which enables techniques based on ODMR. Even better, a number of diamond's material properties conspire to limit the interactions this spin has with the surrounding lattice, making an NV center behave much like an isolated spin. \ifbibtex{\footnote{A number of factors are responsible for the weakness of the interaction of the spin with its surroundings. For example, because the $^{12}\rm{C}$ nucleus has a spin of zero, the ``spin-spin'' interactions are small. Also, the low mass and stiff interatomic bonds of the carbon lattice lead to a high Einstein temperature ($T_E = 2300 \rm{K}$), which makes the interaction of the NV center) with the vibrational modes of the surrounding lattice unusually weak at room temperature.}\else \cite{footnote3} \fi This fact leads to long ``coherence times'' -- the time over which the phase of the spin quantum state remains predictable -- a critical requirement for quantum computing and other applications.\cite{wrachtrup2006processing}

\begin{figure}[!htbp]
\includegraphics[width=3.25in]{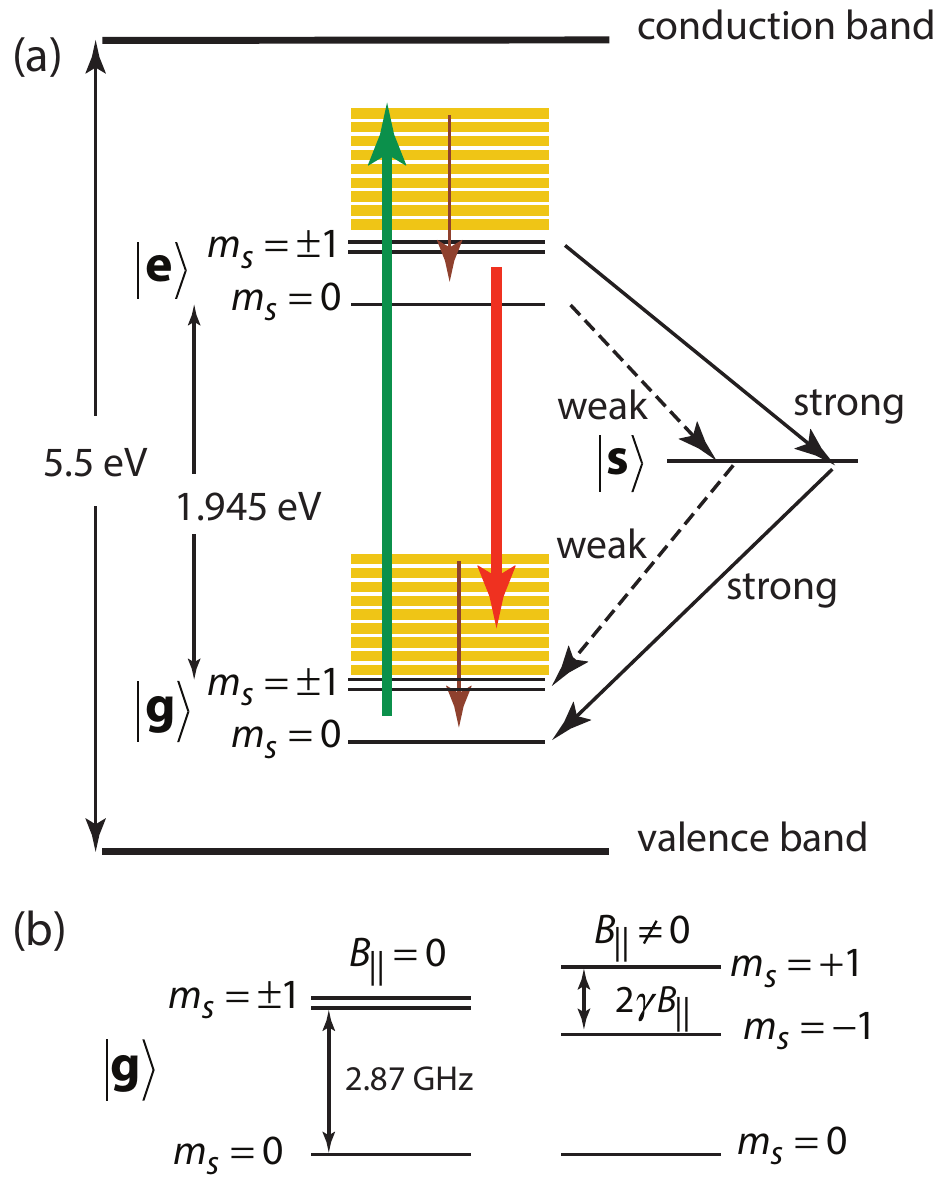}
\caption{Energy levels and transitions in a nitrogen vacancy center. (a) Optical pumping with green light at 532 nm (indicated by the green vertical arrow) induces transitions from the spin-1 ground state $\ket {\bf{g}}$ to the spin-1 excited state $\ket {\bf{e}}$. The yellow bands in the figure represent excited states that have both electronic and vibrational energy. The vibrational energy rapidly dissipates, as indicated by the downward brown arrow. Red fluorescence (red downward arrow)  back to $\ket {\bf{g}}$ is emitted over a range of energies. Again, some energy is transferred to vibrational modes of the system during this process.  Both the excitation and the fluorescence transitions tend to conserve the value of the $m_s$ quantum number. Spin-dependent, non-radiative transitions to and from a long-lived singlet state $\ket {\bf{s}}$ lead to a preferential population of the $m_s = 0$ ground state, producing electron spin polarization of the NV center. Transitions from $\ket {\bf{e}}$ to $\ket {\bf{s}}$ are more likely when the center is in the $m_s = \pm 1$ states, leading to a reduction in the average fluorescence intensity relative to when the center is in the $m_s=0$ state. (b) In the absence of an external magnetic field, the $m_s = 0$ substate of $\ket {\bf{g}}$ is lower in energy than the nearly degenerate $m_s = \pm1$ substates. The relative populations of these substates can be changed if the system is driven with resonant (2.87 GHz) microwave radiation. Application of a magnetic field $B_{||}$ aligned along the NV axis lifts the degeneracy of the $m_s = \pm 1 $ states, giving rise to two distinct resonances separated in frequency by an amount $2\gamma B_{||}$, where $\gamma = 2.8\;\rm{GHz/T}$.}
\label{fig:energylevel3}
\end{figure}

The energy level diagram shown in Fig.~\ref{fig:energylevel3} is a good starting place for understanding the optical properties of an NV center. The system has an excited electronic state $\ket {\bf{e}}$ that lies 1.945~eV above the ground state $\ket {\bf{g}}$. As seen in Fig.~\ref{fig:energylevel3}b, in the absence of an external magnetic field, the $m_s = 0$ substate of $\ket {\bf{g}}$ is lower in energy (by an amount corresponding to a frequency of 2.87 GHz) than the nearly degenerate $m_s = \pm1$ substates. In the presence of an external static magnetic field $\mathbf{B}_0$ there will be a Zeeman shift of the NV frequency associated with the spin states given by

\begin{equation}
\Delta \nu = m_s \cdot \frac{g\mu_B}{h} \cdot B_{||}
 \label{eq:dnu}
\end{equation}

\noindent where $g$ is the $g$-factor for the NV center ground state, $\mu_B$ is the Bohr magneton, and $B_{||}$ is the component of $\mathbf{B}_0$ along the direction of the ``axis'' of the NV center -- the line that connects the nitrogen atom to the vacancy. Therefore the  $m_s = 0$ substate is unshifted and the frequencies associated with the  $m_s = \pm1$ substates, relative to the $m_s = 0$ substate, can be written as 

\begin{equation}
 \nu_\pm = D \pm \cdot \gamma \cdot B_{||}
 \label{eq:gyro}
\end{equation}

\noindent where  $D=2.87\; \rm{GHz}$ and  $\gamma =  g\mu_B/h \approx 2.8 \; \rm{MHz/G} = 28 \; \rm{GHz/T}$  are experimentally measured values for the NV center ground state\ \ifbibtex{\footnote{Eq. \ref{eq:gyro} is only valid in the limit that the magnetic field strength is much smaller than the zero-field splitting between the $m_s=0$ state and the $m_s=\pm1$ states; i.e when $B_{||} << 2.87 \rm{GHz}/\gamma$. This condition is satisfied for all the experiments described here.} \else \cite{footnote4} \fi (See Fig.~\ref{fig:energylevel3}b.) We show in Sec. \ref{Results} that Zeeman shifts will have a noticeable effect on our ODMR spectra for magnetic fields as small as a few tenths of a mT. This is, admittedly, not so impressive when compared to experiments on single NV centers that can detect magnetic fields as small as 3~nT with nanometer spatial resolution!\cite{maze2008nanoscale, balasubramanian2008nanoscale}

Green light (wavelength 532~nm, photon energy $\sim 2.4\;\rm{eV}$) can induce a transition from the ground state $\ket {\bf{g}}$ to the excited state $\ket {\bf{e}}$, which contains 1.945~eV of electronic energy, along with the energy excess transferred to vibrational motion of the NV center. (The yellow bands in  Fig.~\ref{fig:energylevel3} represent states that have both electronic and vibrational energy.) The vibrational energy is rapidly transferred to the surrounding lattice, as indicated by the vertical brown arrow.  Red fluorescence (indicated by the red downward arrow)  is then emitted during transitions back to $\ket {\bf{g}}$. This fluorescence contains photons with a wide range of energies (corresponding to wavelengths in the 600--750~nm band), since again some energy is transferred to vibrational modes of the system. The excited electronic state has a longer lifetime, $\sim 13\;\rm{ns}$, but this is still short enough so that, in the presence of sufficiently intense excitation with green light -- easily obtained with a focused laser -- an NV center can go through many excitation/emission cycles in a short amount of time. For example, for intensities present at the focal point of a 1 mW laser beam, a \textit{single center} can emit $\sim 10^7$ red photons per second. This makes it relatively easy to detect fluorescence from a single center. For the samples used in the experiments described here, in which emission is collected from a large number of  NV centers, the red fluorescence is very bright and is easily visible to the naked eye when viewed through a filter that blocks the green excitation light.

\begin{figure}[!htbp]
\includegraphics[width=3.5in]{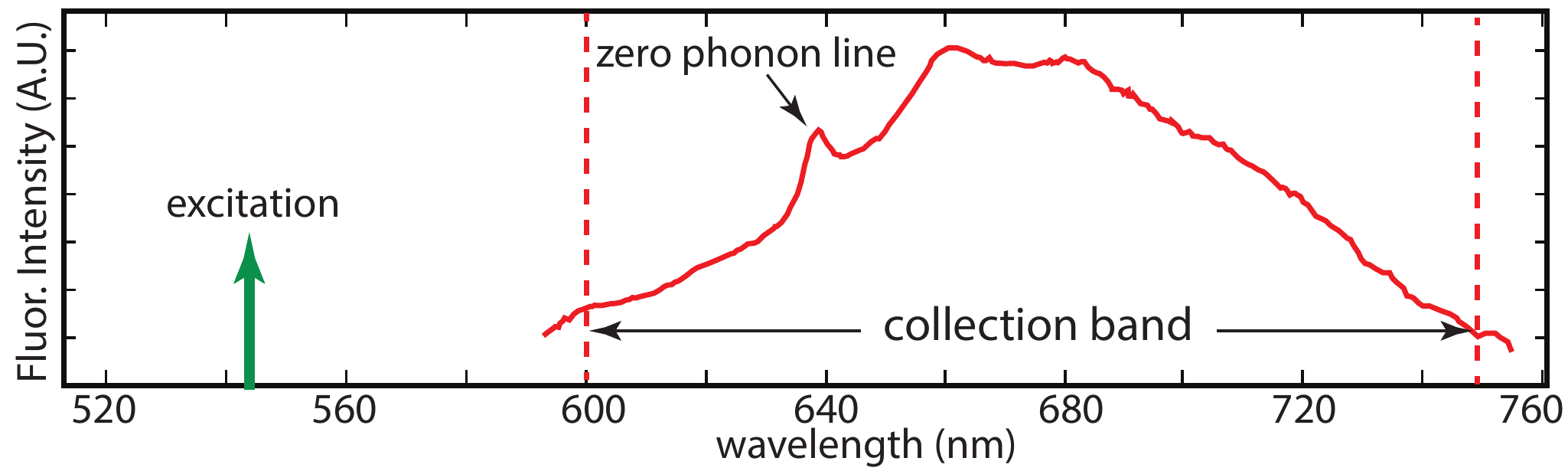}
\caption{Fluorescence spectrum from a collection of nanocrystals with a high concentration of NV centers. Excitation in this case was provided by a green HeNe laser (wavelength 543~nm). In our ODMR experiments, edge filters are used to limit the range of detected fluorescence to a band between 600~nm and 750~nm.}
\label{fig:NVPL}
\end{figure}

Figure~\ref{fig:NVPL} shows a fluorescence spectrum from diamond with a high concentration of NV centers. Note that the spectrum exhibits a peak at a wavelength of 637 nm, which corresponds to photons with an energy of 1.945 eV. This ``zero phonon line'' corresponds to electronic transitions from the excited to ground state of an NV center in which all of the energy difference is carried away by the emitted photon. Such transitions are relatively rare compared to phonon-assisted transitions, in which some energy is also exchanged with the vibrational degrees of freedom of the lattice. The broad fluorescence in the wavelength range 600--750 nm is primarily due to these phonon assisted transitions.

\subsection{Optically detected magnetic resonance}
The intense practical interest in the NV center stems from the observation that its spin state can be both monitored and manipulated using ODMR.  To understand how this is accomplished we need to look more carefully at the energy level diagram of Fig.~\ref{fig:energylevel3}. Local magnetic interactions split the spin-1 (or ``triplet'') ground state $\ket {\bf{g}}$. With the NV axis serving as the reference direction, the $m_s = 0$ state lies $0.018 \;\rm{meV} = 2.87\;\rm{GHz}$ below the nearly degenerate $m_s = \pm 1$ states. At room temperature this energy splitting is small compared to the thermal energy $k_BT$ , so we expect that when the sample is in thermal equilibrium at 300~K these substates should be nearly equally populated. However, in an ODMR experiment the relative populations of these spin states can be altered and measured via interactions with visible light and resonantly tuned microwaves. The basic mechanism can be summarized as follows:\cite{doherty2013nitrogen}

\begin{itemize}

\item Green photons induce phonon-assisted transitions from the spin-1 ground state $\ket {\bf{g}}$ to the spin-1 excited state $\ket {\bf{e}}$, as indicated by the green upward arrow in Fig.~\ref{fig:energylevel3}.

\item Phonon assisted fluorescence due to transitions back to $\ket {\bf{g}}$ is detected in 600--750 nm range. These optical transitions are indicated by the red downward arrow in Fig.~\ref{fig:energylevel3}.

\item  The optical transitions described in the above bullet points tend to preserve spin orientation. (That is, the value of $m_s$ is typically not altered during these transitions.)

\item There also exists a long-lived spin-0 (singlet) state $\ket {\bf{s}}$ that can only be accessed via non-radiative transitions. The selection rules associated with these non-radiative transitions are such that illumination with green laser light leads to a preferential population of the $m_s = 0$ ground state.

\item Transitions from $\ket {\bf{e}}$ to $\ket {\bf{s}}$ are more likely when the center is in a $m_s = \pm 1$ state relative to when it is in a  $m_s = 0$ state. So a center in a $m_s = \pm 1$ state ends up spending a significant fraction of time ``stuck'' in the $\ket {\bf{s}}$  state, leading to a reduction in the average fluorescence intensity relative to when the center is in the $m_s=0$ state.

\end{itemize}

The upshot is that when illuminated with green light, there exists an optical pumping mechanism that tends to drive the NV center into the $m_s = 0$ substate. Also, when an NV center is in a $m_s = \pm 1$ substate the intensity of the red fluorescence will be noticeably less than when it is in the $m_s = 0$ substate.

The relative populations of the $m_s = 0$ and $m_s = \pm 1$ substates can be changed if the system is driven with resonant microwave radiation. In a typical ODMR measurement, the sample is illuminated with green light and the intensity of the red fluorescence is monitored as an applied microwave field is slowly tuned into resonance. At resonance there will be an easily detectable  reduction of fluorescence intensity. 

\section{Experimental Methods}
\subsection{Samples} We made measurements on diamond crystals of three different size scales: large single crystals ($\sim 1- 5\; \rm{mm}$ across), microcrystals ($\sim 10 - 20\; \mu \rm{m}$ across), and nanocrystals ($\sim100 \; \rm{nm}$ across). In all three cases the samples have high concentrations of NV centers, so when we recorded ODMR spectra large ensembles of NV centers were being measured at once.

\textit{Large single crystal diamond.} Large single crystal diamond samples (lateral dimensions up to $\sim5\;\rm{mm}$) with a high concentration of NV centers were provided to us by our collaborators at Element Six, a leading commercial producer of synthetic diamonds.  Diamond single crystals were synthesized at high pressures and high temperatures. These samples typically have an optically smooth [111] surface. As grown, these crystals contained relatively high concentrations of isolated nitrogen impurities ($\sim$ 100 ppm), resulting in color centers that gave them a yellowish-brown hue. Samples were then irradiated with 4.5~MeV electrons (dose $2 \times 10^{18}\rm{e/cm^{2}}$), administered over the course of 2 hours, which created a high concentration of vacancies. Finally, the samples were annealed under vacuum condition ($\approx 10^{-4}\;\rm{Pa}$) for several hours at 800 $^\circ \rm{C}$. At this elevated temperature the vacancies became mobile, resulting in the formation of a high concentration of NV centers. 

\textit{Diamond microcrystals.} Large single crystals with high NV concentrations like the one described above are, to our knowledge, not commercially available. \ifbibtex{\footnote{Element Six sells HPHT diamond with suitably high nitrogen concentrations. See reference \onlinecite{2017arXiv170205332F} for more detailed guidance on creating high NV concentrations in such samples via electron irradiation and annealing.}\else \cite{footnote5} \fi Fortunately there exists a source of single crystal diamond with high NV concentration that is readily available: ``fluorescent microdiamonds'' (Ad\'{a}mas Nanotechnologies MDNV15umHi50mg). To make ODMR measurements on the microcrystals we first deposited a thin layer of a mounting medium (ThermoFisher ProLong Gold, designed for use in fluorescence microscopes) onto a glass coverslip, then used a dry glass pipette to transfer a small number of microcrystals onto the mountant, which we allowed to dry overnight. The orientations of the crystal axes of the microcrystals were randomly distributed. Using the fluorescence microscope described below we could easily isolate the bright fluorescence from a single microcrystal, enabling us to make ODMR measurements on a sample with a well-defined (but unknown) crystal orientation.

\textit{Diamond nanocrystals.} We also performed ODMR measurements on ensembles of 100 nm diameter nanocrystals (Ad\'{a}mas Nanotechnologies ND-NV-100nm). Each nanocrystal contains $\sim$ 500 NV centers. Diamond nanocrystals such as these are non-toxic and can be inserted into various biological systems for \textit{in vivo} magnetometry with high spatial resolution. \cite{balasubramanian2008nanoscale, kucsko2013nanometre, schirhagl2013nitrogen, le2013optical}

The nanocrystals come suspended in de-ionized water (1 mg nanocrystals per a mL water). We prepared samples for ODMR measurements simply by placing a drop of this suspension on a glass coverslip and waiting for the water to evaporate. The nanocrystals adhered naturally to the coverslip, without need for any adhesive. The highest concentration of nanocrystals formed at the outer edges of the drop, much like a coffee stain.

\subsection{Fluorescence microscope} 
\begin{figure}[!htbp]
\includegraphics[width=2.5in]{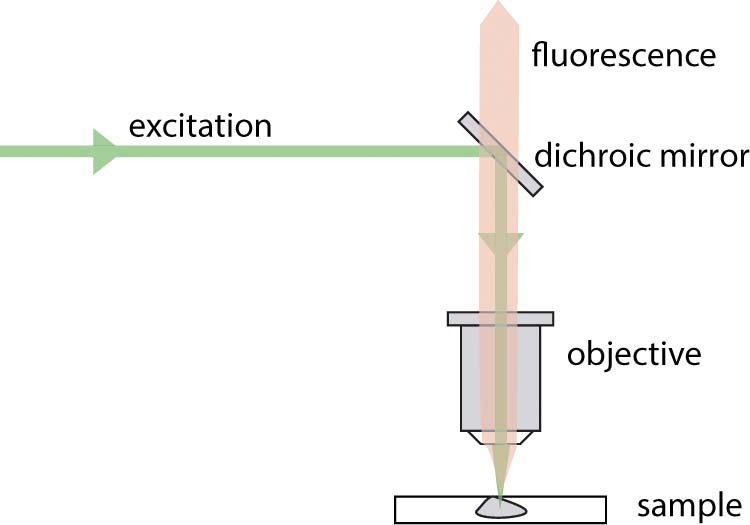}
\caption{Epifluorescence configuration. A dichroic mirror, which reflects green light but transmits red light, allows for collection of the fluorescence signal while rejecting the excitation light.}
\label{fig:epifluor}
\end{figure}

Fluorescence microscopes are widely used in many labs around the world, particularly in the life sciences. Like most commercial instruments, our ODMR setup uses an epifluorescence configuration, shown in Fig.~\ref{fig:epifluor}, in which the same lens is used both to focus the exciting light and to collect the induced fluorescence. A dichroic mirror that is highly reflective to the green excitation light but transparent to the red fluorescence is a critical element in the design.  We excited and detected this red fluorescence from an ensemble of NV centers contained in a relatively small sample volume, determined by how tightly we focused the green laser excitation, typically $\sim 10\; \mu \rm{m}$ in diameter.

\begin{figure}[!htbp]
\includegraphics[width=3.0in]{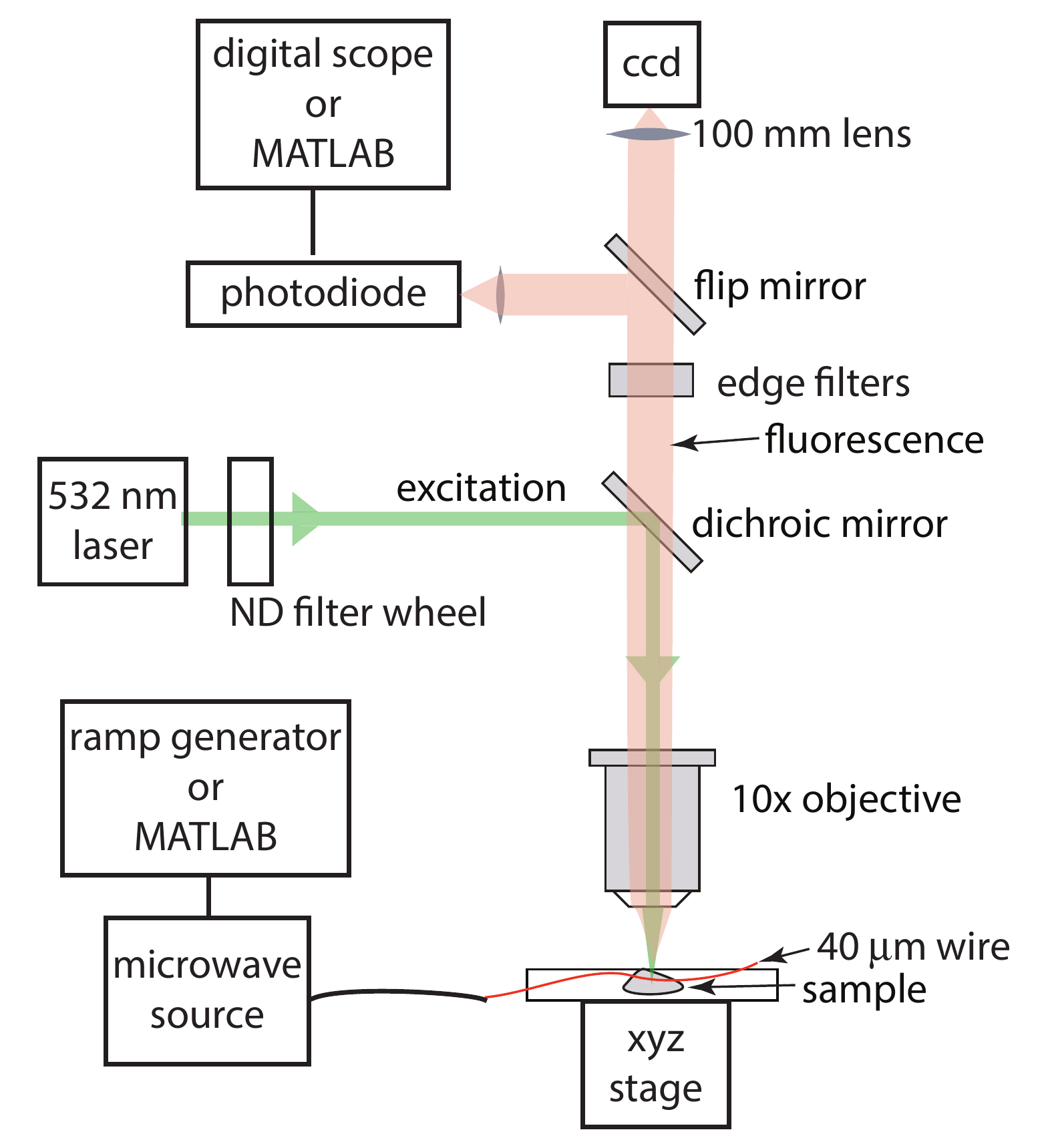}
\caption{Schematic drawing of the fluorescence microscope used for ODMR experiments.}
\label{fig:schematic}
\end{figure}

A detailed schematic drawing of our setup is shown in Fig.~\ref{fig:schematic}. The NV fluorescence was excited by up to 40~mW of green light (wavelength 532 nm) provided by a diode pumped solid state laser (Thorlabs DJ532-40). Since the ODMR technique is dependent on detecting relatively small ($\sim 5 - 10 \%$) changes in the fluorescence intensity, it is important that the laser output power be stable, ideally to better than 1\%. We achieved this stability by powering the laser with a 330 mA constant current source (Thorlabs LDC210C) and mounting it so that its temperature is actively stabilized (Thorlabs TED200C temperature controller with LDM21 mount).  \ifbibtex{\footnote{Excellent lower cost alternative laser systems are the Coherent Compass 215M or the Laserglow Technolgies LCS-532 series. There are even less expensive 532 nm lasers, but these are typically not temperature stabilized. This results in significant fluctuations in output power that may obscure the relatively small variations in fluorescence intensity that we observe here. To reduce the detrimental effects of power fluctuations it helps to record the spectra quickly. Also, for the laser intensities used here, the fluorescence is linearly proportional to the laser power and for longer scans we have had good success in minimizing the effect of laser power fluctuations by monitoring the laser power and normalizing.}\else \cite{footnote6} \fi We placed a filter wheel containing a variety of absorptive neutral density filters (Thorlabs FW1AND) immediately in front of the laser, which allowed us to vary the amount of laser incident on the sample. This also allowed students to work with lower beam intensities during the alignment process. (They were of course also wearing laser safety glasses when working on the alignment. We used glasses that have at least a 2.0 optical density at 532~nm.) A long-pass dichroic mirror with 550~nm cutoff wavelength (Thorlabs DMLP550) reflected the laser light and a 10$\times$ microscope objective (e.g., Thorlabs RMS10X) focused it onto the samples. The sample was mounted on an \textit{xyz} translational stage that allowed us to adjust both the lateral position and the diameter of the focal spot.

The same 10$\times$ microscope objective collected and collimated the fluorescence, which was transmitted through the dichroic mirror. A fair amount of elastically scattered green light was also collected by the microscope objective, but this was mostly rejected by the dichroic mirror. A pair of edge filters was also used: An additional long pass filter with a cutoff of 600~nm (Thorlabs FEL0600) further attenuated the green light, while a short pass filter (Thorlabs FES0750) attenuated light with wavelengths longer than 750~nm, so that the detected light is limited to the band associated with the NV fluorescence (see Fig.~\ref{fig:NVPL}. The fluorescent light was detected by an adjustable-gain photodiode (Thorlabs PDA36A), whose output was recorded  either with a digital oscilloscope (Tektronix 2014B) or with a low-noise digital multimeter (Keysight 34460A) connected to a computer under the control of a MATLAB program.

\begin{figure}[!htbp]
\includegraphics[width=3.25 in]{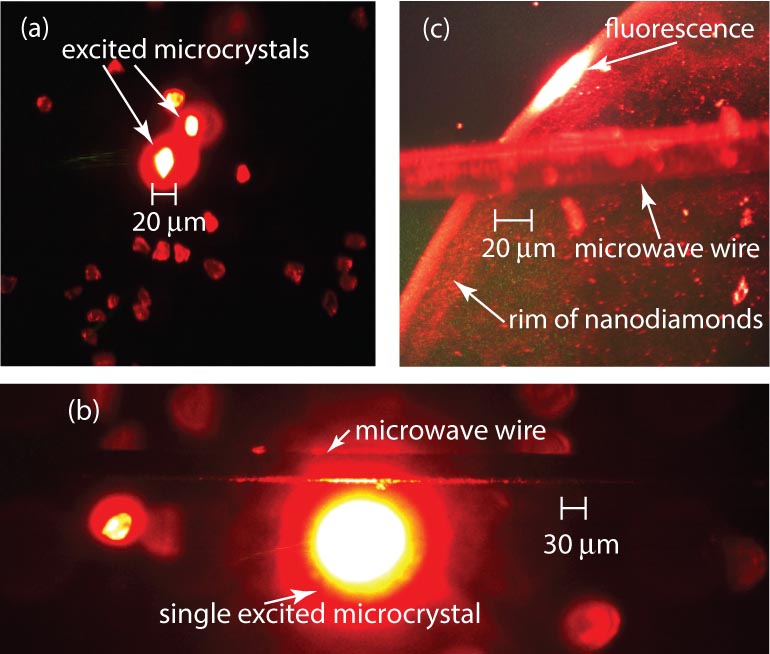}
\caption{Image of a diamond microcrystal sample under laser excitation, recorded by the  camera. (a) Microcrystals (average diameter $15\;\mu\rm{m}$) with laser loosely focused to a diameter of about $20\;\mu\rm{m}$, exciting a bright red fluorescence in two adjacent microcrystals. (b) Laser focused on a single microcrystal located adjacent to the thin wire, where the microwave field intensity is greatest. (c) Diamond nanocrystal sample with laser focused at the rim of the sample, where the concentration of nanocrystals is highest.}
\label{fig:nanocystalccd}
\end{figure}

To aid in focusing the microscope, the mirror that directs the fluorescence onto the photodiode could be flipped out of position so that a CMOS camera (Thorlabs DCC1645C) outfitted with a 100~mm lens (Thorlabs MVL100M23) could image the collected light. Since a 10$times$ microscope objective has a focal length of 16~mm, this made images formed on the camera's sensor $\frac{{100\,{\rm{mm}}}}{{16\,{\rm{mm}}}} = 6$ times larger than the object. The images were viewed on a computer monitor using software that allows for additional ``digital zooming.'' Typical images are shown in Fig.~\ref{fig:nanocystalccd}. Note that everything is red because the image is viewed through dichroic and long-pass filters. For the image in Fig.~\ref{fig:nanocystalccd}(a) the excitation laser was loosely focused to a diameter of about $20\;\mu\rm{m}$, which excited a bright red fluorescence in two adjacent microcrystals (average diameter $15\;\mu\rm{m}$). In Fig.~\ref{fig:nanocystalccd}(b) the laser was focused on a single microcrystal located close to a thin wire through which microwave frequency currents flow, since this is where the microwave field intensity was greatest. (Note that the bright fluorescence saturates the image, making the microcrystal appear much larger than it actually is.) Fig.~\ref{fig:nanocystalccd}(c) shows a nanocrystal sample with the laser focused at the rim of the sample, where the concentration of nanocrystals was highest. The laser focus was again chosen to be close to the microwave carrying wire.

\begin{figure}[!htbp]
\includegraphics[width=3.25 in]{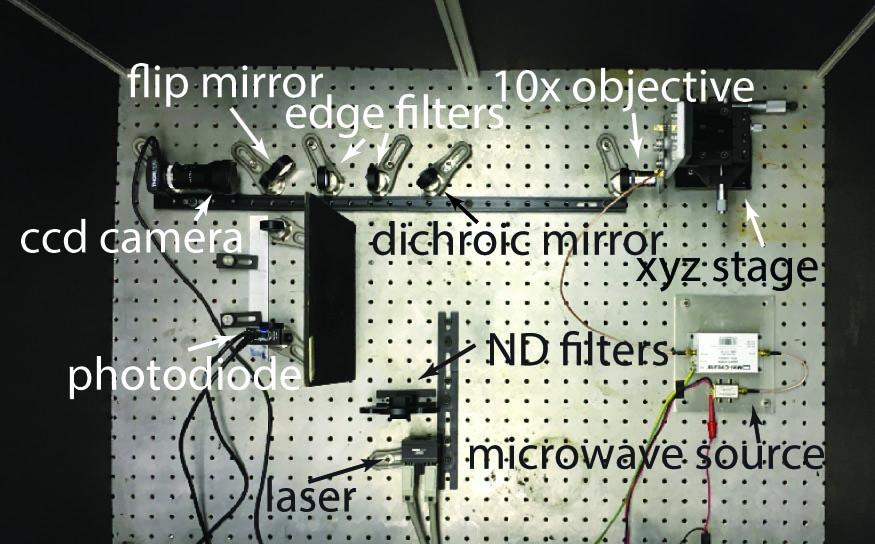}
\caption{Inside the black box. We believe that activities where students build their own instruments helps improve both their engagement and their learning.}
\label{fig:blackbox}
\end{figure}

A photo of the optical setup is shown in Fig.~\ref{fig:blackbox}. The arrangement is simple enough so that it can be constructed by novices starting from an empty optical breadboard in an hour or two. The cost of the setup is also relatively modest, in the two to three thousand dollar range. Alignment is simplified by mounting all the optical elements so that their centers are at a common height. (Specifically, we use a holder for 1-inch optics (Thorlabs LRM1) mounted on top of a 1-inch diameter pedestal post (Thorlabs RS1P8E) and held in place on the optical breadboard with a clamping fork (Thorlabs CF125C). The pedestal posts can be quickly positioned and fixed in position along a straight line by sliding against an an aluminum bar clamped to the optical breadboard.) The entire setup is enclosed in a quasi-light tight box constructed of black foamboard panels that slide into construction rails (Thorlabs XE25). In our experience this ``build your own'' aspect of the activity greatly increases student engagement and learning.
 
\subsection{Microwave source} To record ODMR spectra we added a microwave field with tunable frequency in a $\pm 0.2\;\rm{GHz}$ band around the 2.87 GHz resonance. The microwaves were broadcast from a 40~\;$\mu \rm{m}$ diameter copper wire positioned close to the region from which the NV fluorescence was collected.  (See Fig.~\ref{fig:nanocystalccd}.) The ends of the wire, which is about 5~mm long, are soldered to surface mount pads on a small custom-made printed circuit board, which connects the pads to a SMA jack via short traces.  The magnetic field component of the microwaves is proportional to the current flowing through the wire $I_{rf}$ and inversely proportional to the distance from the wire $r$:

\[{B_{rf}} = \frac{{{\mu _0}{I_{rf}}}}{{2\pi r}}\]

\noindent By using a small diameter wire and focusing near the wire, we were able to obtain sufficiently intense fields using microwave sources that deliver only modest amounts of power. In our experiments we used a microwave synthesizer that can deliver up to 20~dBm = 0.1 W of average power, which proved more than enough to observe a clear ODMR signal. ODMR spectra were recorded by monitoring the fluorescence intensity as the microwave frequency was varied in the frequency range 2.7--3.1 GHz.

\begin{figure}[!htbp]
\includegraphics[width=3.0in]{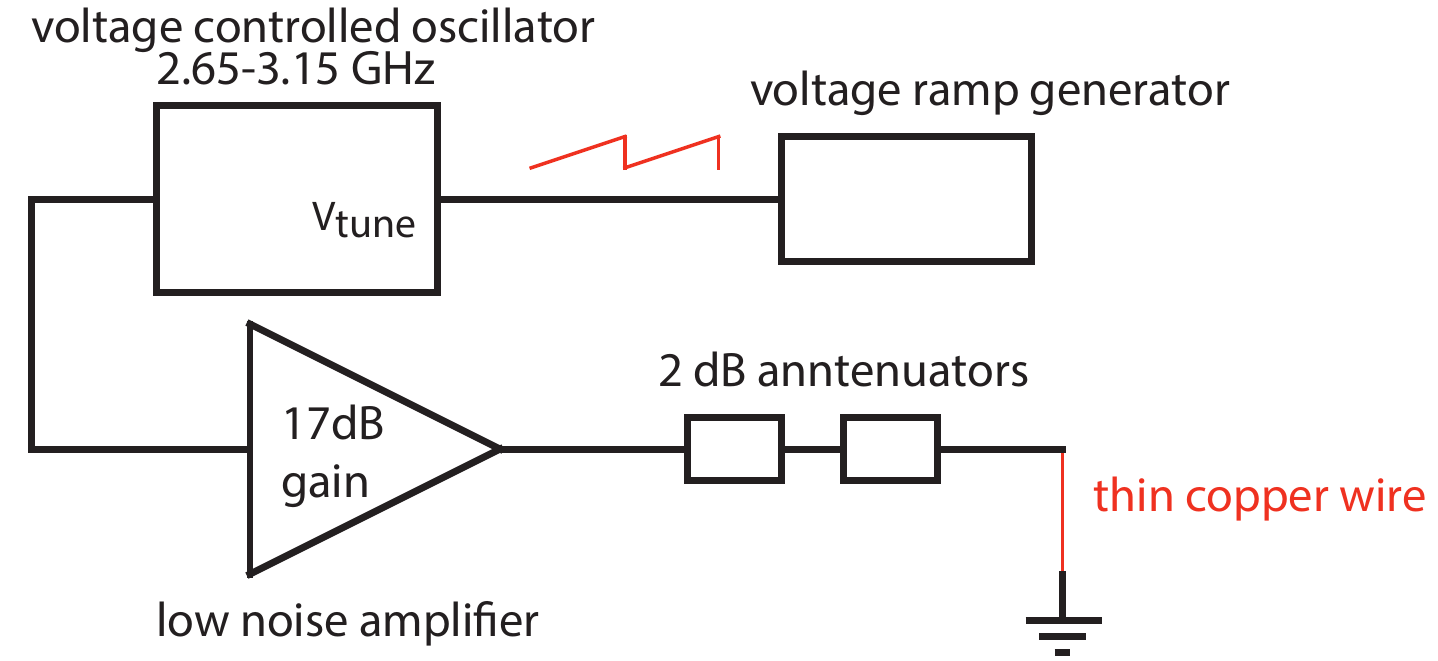}
\caption{A simple microwave source. A voltage controlled oscillator supplies a microwave signal whose frequency can be varied by adjusting a dc tuning voltage supplied by a ramp generator. This signal is boosted by a low noise microwave amplifier, which is capable of delivering over 100 mW of power into a 50 $\Omega$ load.}
\label{fig:MWS}
\end{figure}

We experimented with two different custom-built microwave sources. One design, which is also suitable for more sophisticated pulsed ODMR measurements, is digitally controlled and uses a phase locked loop to achieve a frequency stability and accuracy of less than 1 MHz with very low phase noise. But, since the ODMR resonances reported below have widths that are large compared to 1 MHz, we were also able to use the simpler lower resolution microwave source shown in Fig.~\ref{fig:MWS}. In this design a voltage controlled oscillator (Mini-Circuits ZX95-3150+) supplies a microwave signal whose frequency can be varied by adjusting a dc tuning voltage. This signal can be boosted by a low noise microwave amplifier (Mini-Circuits ZRL-3500), although this amplification may not be necessary when the fluorescence is collected from a small volume very close to the thin wire.

We used a microwave spectrum analyzer to determine that the frequency vs. tuning voltage behavior of this circuit is well described by the relation 

\begin{equation}
\nu = \bigg{(}0.068\,\frac{{{\rm{GHz}}}}{{\rm{V}}}\bigg{)} {V_{tuning}} + 2.476\,{\rm{GHz}}
 \label{eq:nutune}
\end{equation}

\noindent in the frequency range $2.7\,{\rm{GHz}} < \nu < 3.1\,\,{\rm{GHz}}$. The microwave frequency was swept linearly over time by using a sawtooth signal from a function generator (Agilent 33210A) to control the tuning voltage. We observed that the spectral width of the output was less than 10~MHz, which was perfectly adequate for our purposes here.

The microwave power could be incrementally varied by placing a number of 2 dB attenuators (Omni Spectra 2082-6171-02) in series with the load. We have not designed the load to have an impedance equal to the 50~$\Omega$ output impedance of the microwave source, so reflections from the mismatched load could possibly interfere with the operation of the amplifier. Having at least one or two of the 2~dB attenuators  between the source and the load helps reduce any potential problems caused by these reflections -- the attenuators act on microwaves both ``coming and going'' -- while still delivering adequate microwave power. Another option is to insert an isolator, such as a TRAK 60A301, in series with the load, which acts to prevent reflected waves from traveling back to the amplifier.

\subsection{Zeeman effect}
We expect that Zeeman splittings of the $m_s = \pm1$ substates will occur in the presence of a static magnetic field.  To investigate this, we used a 20-turn coil wrapped around the 10$\times$ microscope objective and driven by a 0--2~A current source to create the desired static fields. This geometry led to magnetic fields directed along the optical axis of the microscope. The field strength at the objective varied from 0~mT to 1.2~mT, based on calculation and corroborated by measurements made with a Hall sensor. We also created magnetic fields of different strengths by placing a strong permanent magnet on a translational stage and moving it relative to the sample.

\subsection{A simpler setup}

\begin{figure}[!htbp]
\includegraphics[width=3.5in]{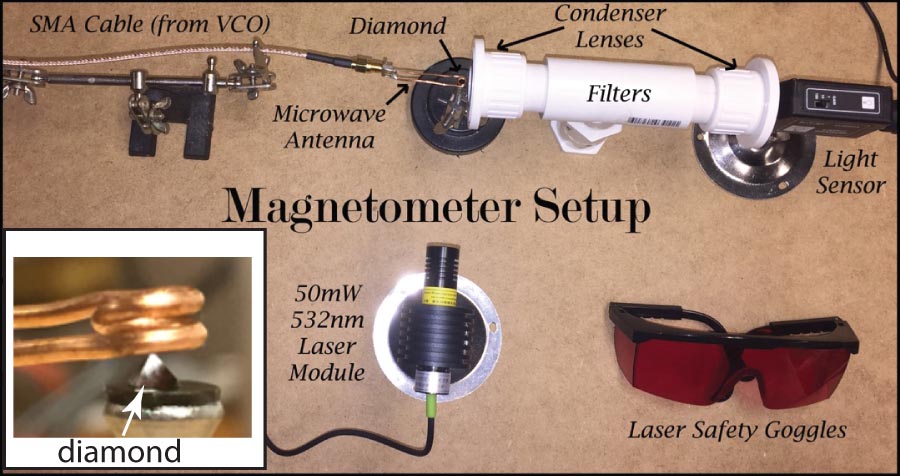}
\caption{A simplified ODMR setup. The elements were all held in homemade mounts, which can be quickly arranged on a table top. A coiled wire, shown in the inset, produced a relatively uniform microwave field throughout the sample.}
\label{fig:simplesetup}
\end{figure}

Figure~\ref{fig:simplesetup} shows a simplified version of an ODMR experiment that is significantly simpler to set up than the  fluorescence microscope described above. A circuit similar to the one shown in Fig.~\ref{fig:MWS} supplies microwave currents to a coiled wire that is positioned above the sample (see the inset of Fig.~\ref{fig:simplesetup}). This geometry allowed us to illuminate the sample from the side with an unfocused laser beam, exciting a large volume within the sample. A pair of simple lenses (Thorlabs ACL25416U-B) collimate the fluorescence, which is then passed through edge filters before being focused onto a light sensor (Pasco CI-6604). The elements are all held in homemade mounts, fashioned from PVC pipe, which can be quickly arranged on an ordinary table top. Not only is this version easier to align compared to the fluorescence microscope described above, but the cost is also significantly lower, since there is no longer a need for a microscope objective,  camera, expensive optical mounts or an optical breadboard. The laser  is an inexpensive (\$50) model purchased online from NewGazer Tech. This laser is not temperature controlled and therefore exhibits large fluctuations (about 25\%) in power output. The characteristic time scale of these drifts is on the order of a few seconds or longer, so this is not a big problem provided the scan times are kept sufficiently small. The only catch is that this approach only works for relatively large diamond crystals (lateral dimensions at least 1~mm), with high concentrations of NV centers, which, unfortunately, are not commercially available.

\section{Results} \label{Results}
When illuminated with a few milliwatts of green laser light, the red fluorescence from all of the samples was bright enough to be easily visible to the naked eye when viewed through a long pass filter. In all of our ODMR spectra we observed a distinct decrease in the intensity of the red fluorescence when the frequency of the microwave radiation is in the vicinity of 2.87~GHz, which corresponds to when the microwaves are resonant with the splitting between the $m_s = 0$ state and the $m_s = \pm 1$ states (see Fig.~\ref{fig:energylevel3}). This served as an unmistakable signature that NV centers were making a significant contribution to the red fluorescence.

\subsection{ODMR in diamond single crystals}

\begin{figure}[!htbp]
\includegraphics[width=3.0in]{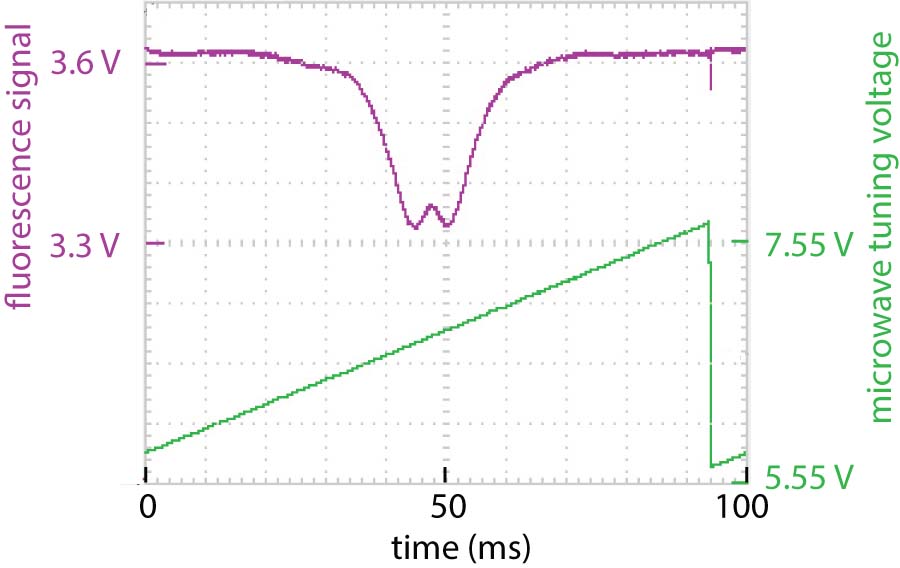}
\caption{An oscilloscope screenshot of an ODMR spectrum recorded from a single microcrystal. The pink trace shows the fluorescence signal, while the green trace shows the microwave tuning voltage.}
\label{fig:microcrystal}
\end{figure}

An oscilloscope screenshot of an ODMR spectrum recorded from a single microcrystal is shown in Fig.~\ref{fig:microcrystal}. The pink trace shows the fluorescence signal while the green trace shows the microwave tuning voltage. When the microwave frequency was tuned to resonance, the fluorescence intensity was about 8\% lower than its off resonance value. Note that near resonance the spectrum consists of a pair of closely spaced dips. Since in this case there is no applied magnetic field, the energy level diagram of Fig.~\ref{fig:energylevel3}b -- as well as Eq.~\ref{eq:gyro} -- suggests that at resonance the spectrum should exhibit only a single dip.  This ``zero field splitting''  has been attributed to the presence of strain that reduces the 3-fold symmetry of the environment surrounding each NV center, resulting in mixing and shifting of the $m_s=\pm1$ levels.\cite{Gruber-Science1997,tisler2009fluorescence}  The result is that, in the presence of strain, the degenerate  $m_s=\pm1$ levels are replaced by a pair of non-degenerate levels whose frequencies relative to the $m_s=0$ level are given by\cite{Gruber-Science1997,acosta2010temperature,acosta2011optical}

\begin{equation}
 \nu _ {\pm } =D \pm \sqrt {{E^2} + {{\left( {\gamma  \cdot {B_{||}}} \right)}^2}}
 \label{eq:nupm}
\end{equation}

\noindent where $E$ is a parameter determined by the magnitude of the zero field splitting between $\nu_+$ and $\nu_-$. Note that the resonant frequencies given by  Eq.~\ref{eq:nupm} approach those given by Eq.~\ref{eq:gyro} in the limit $\gamma  \cdot {B_{||}} >> E$ (i.e., when the Zeeman splitting becomes large compared to the zero field splitting).

\begin{figure}[!htbp]
  \includegraphics[width=3.25in]{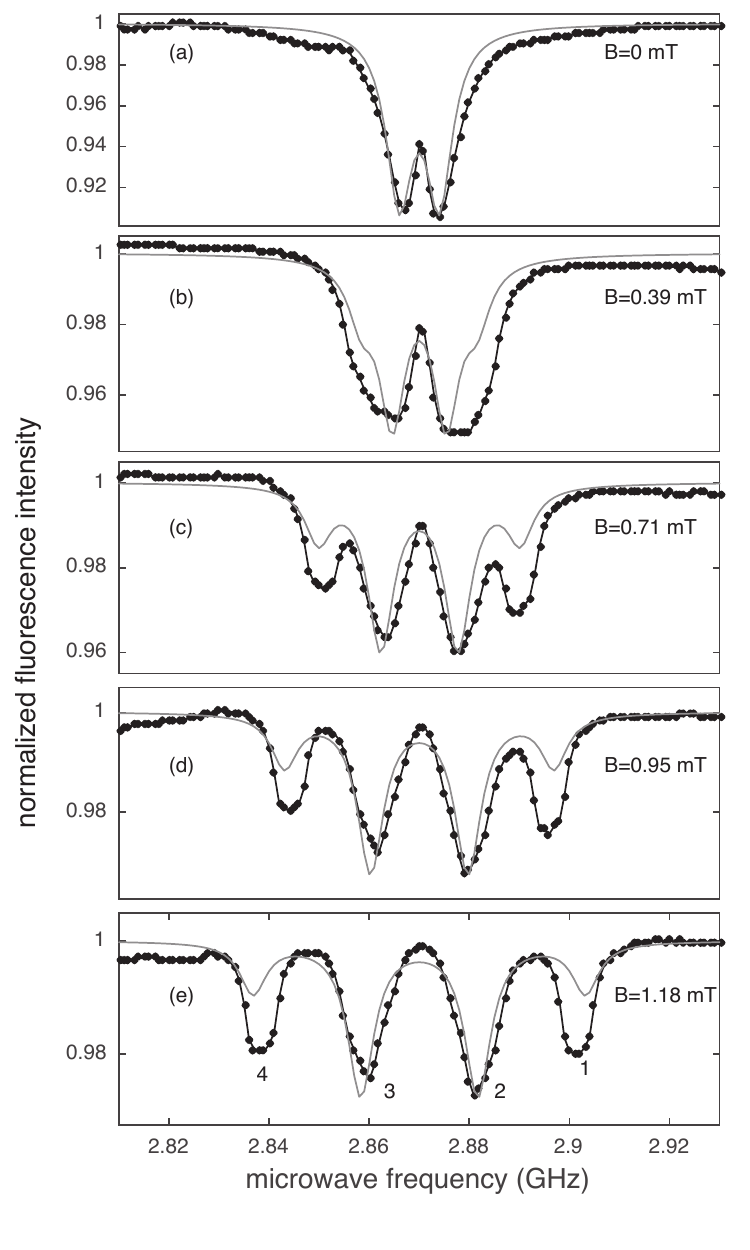}
\caption{ODMR in diamond single crystals: Theory and experiment. Spectra were recorded in the presence of an external magnetic field directed along the [111] direction with magnitude varying from 0~mT to 1.18~mT. Experimental data points are black dots. As the Zeeman shift becomes large compared to the zero field splitting four distinct dips, labelled 1-4, are observed.}
\label{fig:singlecrystal}
\end{figure}

Spectra recorded from a larger single crystal diamond sample are shown in Fig.~\ref{fig:singlecrystal}.  The [111] surface of the crystal was oriented perpendicular to the optical axis of the microscope. Knowing the orientation of this sample will prove useful in the analysis that follows. The spectra are normalized so that when the microwave frequency is far from resonance the fluorescence is scaled to have a value of 1.

The spectrum for the case where there is no applied external magnetic field (Fig.~\ref{fig:singlecrystal}(a) exhibits a reduction in fluorescence intensity $\sim$ 8 \% when the microwave frequency is in the vicinity of 2.87~GHz. Referring to Eq.~\ref{eq:nupm}, we can immediately estimate the values of the three parameters: (1) The  parameter $D$ determines the frequency where the fluorescence decrease is centered, so we estimate $D\approx2.87\;\rm{GHz}$. (2) For this spectrum there is no intentionally applied magnetic field, so we know $B_{||}\approx0$, (3) In zero magnetic field ${\nu _ \pm } = D \pm E$, so the difference between the two transition frequencies is $2E$. This corresponds to the separation between the two nearby minima that appear in the spectrum, which is about 0.1~GHz, so $E\approx0.005\;\rm{GHz}$.

The ODMR spectrum associated with a single transition can best be modeled by a Lorentzian line shape.\cite{Gruber-Science1997} Therefore we can gain more accurate values for these parameters by fitting  the  normalized fluorescence spectrum arising from transitions $\nu_+$ and $\nu_-$ to a function $I\left( \nu  \right) \equiv 1 - f\left( \nu  \right)$, where $f\left( \nu  \right)$ represents the microwave induced decrease in this normalized intensity.: 

\begin{equation}
f\left( \nu  \right) \equiv C\left( {\frac{{{{\left( {\frac{\Gamma }{2}} \right)}^2}}}{{{{\left( {\nu  - {\nu _ + }} \right)}^2} + {{\left( {\frac{\Gamma }{2}} \right)}^2}}} + \frac{{{{\left( {\frac{\Gamma }{2}} \right)}^2}}}{{{{\left( {\nu  - {\nu _ - }} \right)}^2} + {{\left( {\frac{\Gamma }{2}} \right)}^2}}}} \right)
\label{eq:zerofield}
\end{equation}

\begin{figure}[!htbp]
  \includegraphics[width=3.25in]{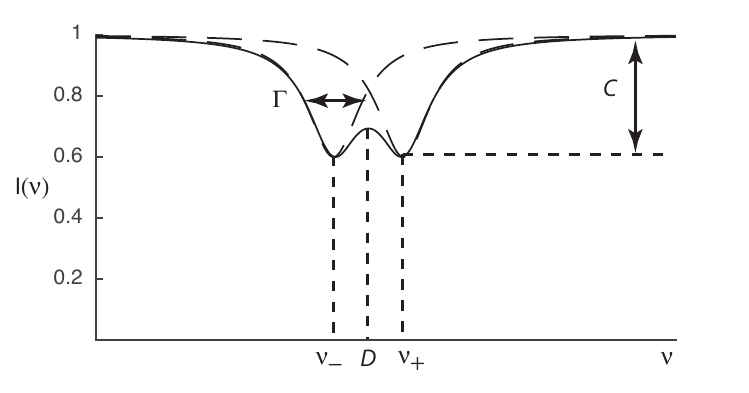}
\caption{The solid curve depicts function $I\left( \nu  \right)$ used to model the normalized ODMR fluorescence spectrum from a pair of transitions at frequencies  $\nu_+$ and $\nu_-$. The dashed curves represent the contributions from each of the transitions. Each transition is assumed to have a Lorentzian line shape with full width at half maximum $\Gamma$. The parameter $C$ determines the contrast.}
\label{fig:doublelorentz}
\end{figure}

\noindent where  $\Gamma$ corresponds to the full width at half maximum of each of the Lorentzian line shapes. A graph of  $I\left( \nu  \right)$ is shown as a solid curve in Fig.~\ref{fig:doublelorentz}. The dashed curves represent the contributions from each of the transitions. The parameter $C$ determines the ``contrast,'' the fractional drop in intensity when the microwaves are tuned to resonance. Adjusting the parameters that appear in Eq.~\ref{eq:zerofield} to obtain a best fit of the spectrum in  Fig.~\ref{fig:singlecrystal}(a)  we find: $C=0.08$, $E=0.0040\;\rm{GHz}$, $D=2.870\;\rm{GHz}$ and $\Gamma=0.0063\;\rm{GHz}$.  

Figure~\ref{fig:singlecrystal}(b)--\ref{fig:singlecrystal}(e) shows OMDR spectra obtained in the presence of a small applied static magnetic field $\mathbf{B}_0$ of various magnitudes, directed perpendicular to the diamond [111] surface. (Choosing the  magnetic field to lie along the [111] direction will make it easier to interpret pattern of Zeeman splitting. But other orientations can certainly be investigated.) As the magnitude of the magnetic field increases, the ODMR spectra gradually transforms to a pattern of four distinct fluorescence dips, labelled 1--4 in Fig.~\ref{fig:singlecrystal}(e). Note that Zeeman shifts are  readily observable using only modest magnetic field strengths produced by small hand-wound coils. In fact, care must be taken to avoid unintentionally introducing magnetic fields of comparable magnitude. For example, it was important to avoid using magnetic mounts for the microscope components, since the stray fields from these mounts can alter the ODMR measurements.

We have also made more qualitative measurements, in which the magnetic field strength was varied by changing the position of  a small permanent magnet in the vicinity of the sample. This gives spectra that transform continuously on the oscilloscope screen as the magnet is moved, which is a wonderful effect.

\begin{figure}[!htbp]
\includegraphics[width=3in]{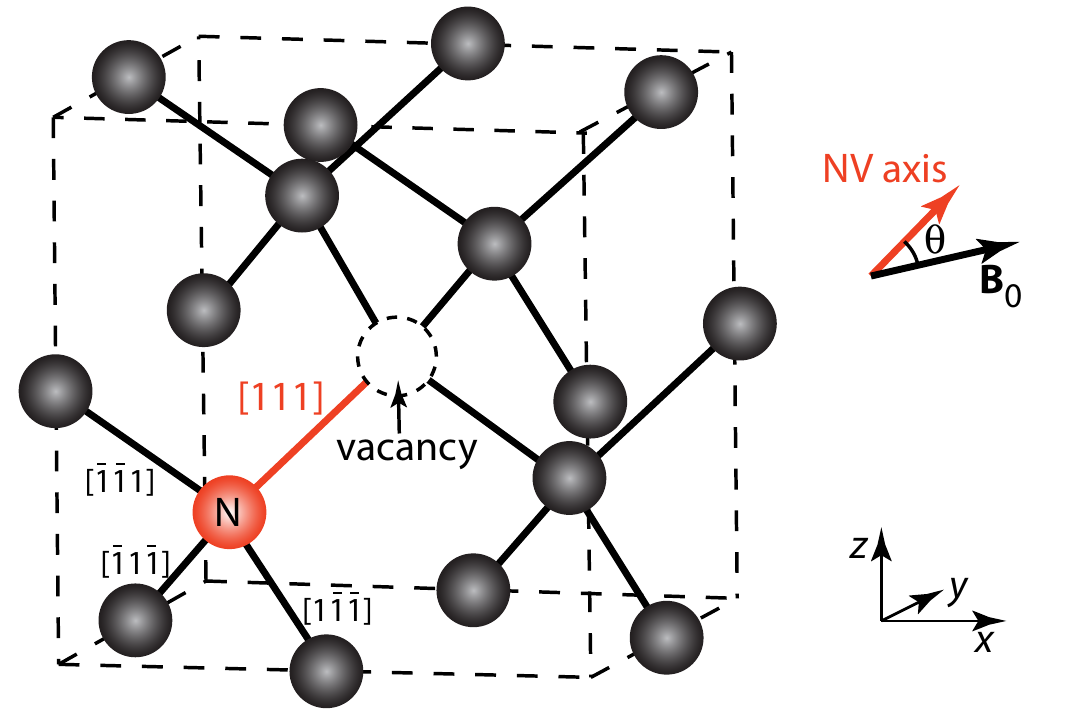}
\caption{NV center with its axis oriented in the [111] direction. The angle between the NV axis and $\mathbf{B}_0$ is denoted by the angle $\theta$. NV centers can also have their axes oriented in the $\left[ {1\,\bar 1\,\bar 1} \right]$, $\left[ {\bar 1\,1\,\bar 1} \right]$, and $\left[ {\bar 1\,\bar 1\, 1} \right]$ directions.}
\label{fig:NVorientations}
\end{figure}

We can explain the main features of these spectra with a simple model. According to Eq.~\ref{eq:nupm}, the Zeeman shift of the ground state energy levels of the NV center depends on  $B_{||}$, the component of the applied magnetic field parallel to the axis of the NV center. Figure~\ref{fig:NVorientations} shows an NV center with its axis oriented in the [111] direction. The angle between the NV axis and $\mathbf{B}_0$ is denoted by the angle $\theta$. NV centers can also have their axes oriented in the $\left[ {1\,\bar 1\,\bar 1} \right]$, $\left[ {\bar 1\,1\,\bar 1} \right]$, and $\left[ {\bar 1\,\bar 1\, 1} \right]$ directions. If we assume that these four orientations are equiprobable, then for $\mathbf{B}_0$ oriented perpendicular to [111] surface, we expect that 1/4 of the NV centers will have $B_{||}=B_0 \cos 0^\circ = B_0$ and 3/4 of the NV centers will have $B_{||}=B_0 \cos 109.5^\circ = -\frac {B_0}{3} $.

Model spectra are given in the gray curves included with  the $B_0\ne0$  spectra  in Fig.~\ref{fig:singlecrystal}~(b)--\ref{fig:singlecrystal}(e). Here we assume that each spectrum is a 3:1 weighted superposition of  two functions of the form $I\left( \nu  \right)$, one calculated using  $B_{||}=B_0$ and  the other  using $B_{||}= -\frac {B_0}{3}$, where the value  of $B_0$ is calculated from the coil geometry and the measured current flowing in the coils. \textit{Thus no new adjustable parameters are needed to fit the $B \ne 0$ spectra.}  This is the basis of using ODMR spectroscopy with NV centers  to measure magnetic fields.\cite{maze2008nanoscale, balasubramanian2008nanoscale, acosta2011optical, glenn2015single} In fact the small discrepancies between the dip locations in the experimental data relative to the model spectra are likely due to the fact that  the model relies on a value of $B_0$ determined indirectly. A value of $B_0$ obtained by fitting the model spectra to the experimental spectra would be a more accurate measure of the magnetic field strength in the region being sampled.

Our analysis suggests that we can identify the four fluorescence dips as corresponding to the following values of $m_s$, $\theta$, and $B_{||}$: 
\vspace{10 pt} 

\hspace{20 pt} $1 \to  {{m_s} = + 1,\;\theta = 0^\circ },\;B_{||}=B_0$

\hspace{20 pt} $2 \to  {{m_s} = -1,\;\theta = 109.5^\circ },\;B_{||}=-B_{0}/3$

\hspace{20 pt} $3 \to {{m_s} = + 1,\;\theta = 109.5^\circ },\;B_{||}=-B_{0}/3$

\hspace{20 pt} $4 \to  {{m_s} = - 1,\;\theta = 0^\circ },\;B_{||}=B_0$

\vspace{10 pt} 

\begin{figure}[!htbp]
\ifeps
  \includegraphics[width=3.25in]{singlecrystal2.eps}
\else
  \includegraphics[width=3.25in]{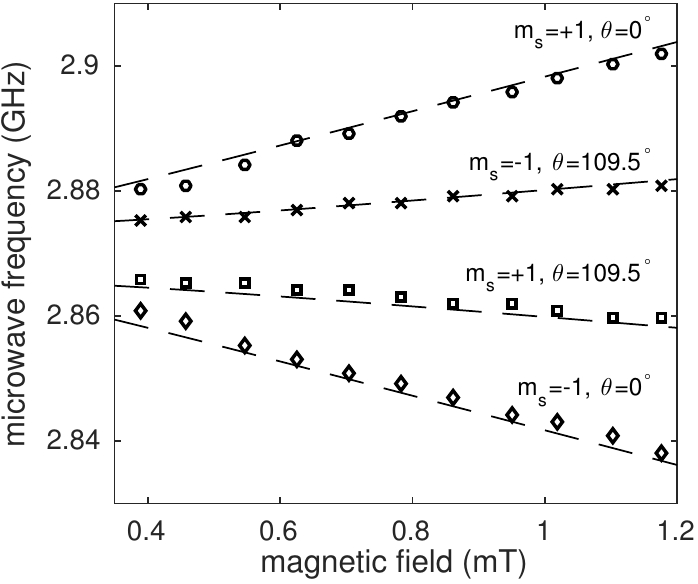}
\fi  
\caption{Zeeman splitting of the NV ground state as a function of magnetic field strength. The magnetic field is directed along the [111] direction. We use different symbols to mark the locations of the different fluorescence dips: $\circ \to \left( {{m_s} = + 1,\theta = {0^\circ }} \right)$, \mbox{\footnotesize $ \times$ } $\to \left( {{m_s} = -1,\theta = 109.5^\circ } \right)$, \mbox{\tiny $ \square$} $\to \left( {{m_s} = +1,\theta = 109.5^\circ } \right)$, \mbox{\footnotesize $ \lozenge$ } $\to \left( {{m_s} = -1,\theta = 0^\circ } \right)$. The dotted lines show the location of these four features predicted by Equation ~\ref{eq:nupm}. }
\label{fig:singlecrystal2}
\end{figure}

\noindent A comparison between the predicted and the experimentally observed dip locations is shown in Fig.~\ref{fig:singlecrystal2}, where we plot the frequency of each of the four dips from Fig. \ref{fig:singlecrystal} as a function of the magnitude of $\mathbf{B}_0$. At low fields the dips are hard to discern -- see, for example, the spectrum in Fig.~\ref{fig:singlecrystal}(b) -- so we plot only the range where distinct dips are observed. The dotted lines, which show the location of these four features predicted by Eq. ~\ref{eq:nupm}, are in good agreement with the experimental observations. Dips 1 and 4 are associated with those NV centers whose axes are along the direction of applied magnetic field, while dips 2 and 3 are associated with centers whose axes make an angle of $109.5^\circ$ with the applied field. Note that for the latter case $B_{||}$ is negative, so Eq.~\ref{eq:dnu} implies that the frequency of the $m_s=+1$ ($m_s=-1$) transition decreases (increases) as the magnetic field strength increases. Since we expect only one out of every four centers to have  $\theta = 109.5^\circ,$ we anticipate that that the amplitudes of dips 1 and 4 should be roughly one third the amplitude of dips 2 and 3. The amplitude ratio in the measured spectra is clearly more than the expected 1:3, for reasons that we cannot explain. 

\subsection{ODMR in diamond nanocrystals}

\begin{figure}[!htbp]
\ifeps
  \includegraphics[width=3.25in]{nanocrystal.eps}
\else
  \includegraphics[width=3.25in]{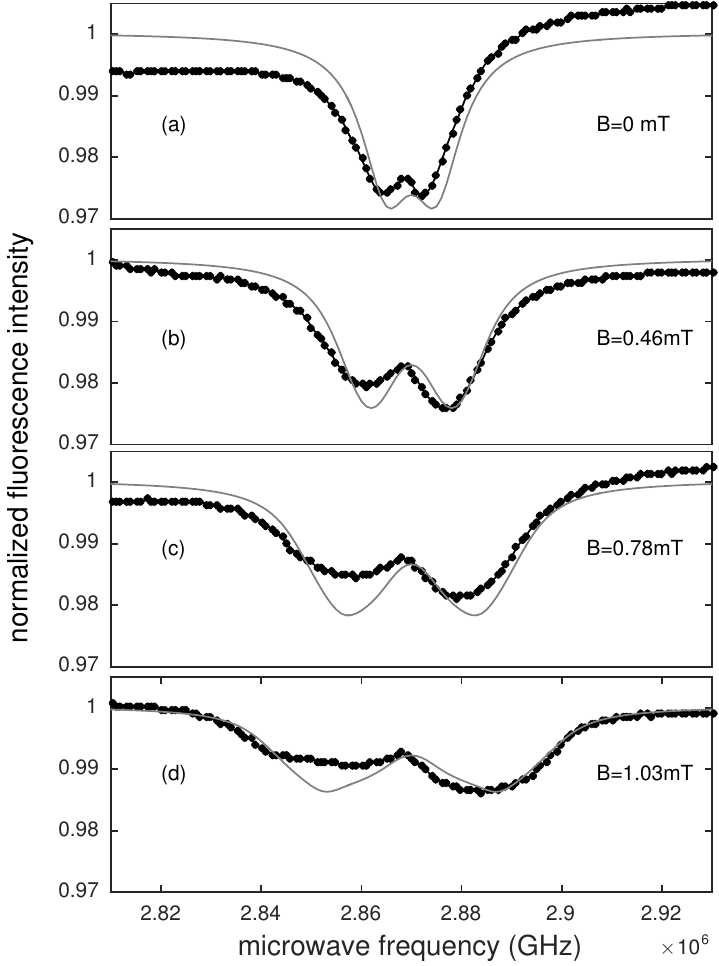}
\fi  
\caption{Zeeman broadening of ODMR in diamond nanocrystals for different magnetic field strengths. Due to the random orientation of the nanocrystals we no longer observed distinct Zeeman levels as the field strength increased, but rather a steady broadening of the $\mathbf{B}_0 = 0$ resonance.}
\label{fig:nanozeeman}
\end{figure}

Optically detected magnetic resonance spectra  recorded from diamond nanocrystals are shown in Fig.~\ref{fig:nanozeeman}. The $\mathbf{B}_0 = 0$ spectrum (Fig.~\ref{fig:nanozeeman}(a)) again exhibits a pair of closely spaced dips in fluorescence intensity in the vicinity of 2.87~GHz. This is similar to what we observed in our single crystal measurements, although the resonance in this case is somewhat broader and has a smaller ($\sim 4\;\%$) contrast with the off-resonance fluorescence. This is likely due to inhomogeneous broadening from sampling an ensemble of nanocrystals with slightly different resonant frequencies.\cite{Gruber-Science1997} The data shown here were collected using relatively long scan times (about 1 minute). During this time there is a small  upward drift in the laser power, which manifests in the spectrum shown in Fig.~\ref{fig:nanozeeman}(a).

We can model these spectra using an approach similar to the one used to analyze the single crystal spectra. As before, we start with the $B_0=0$ spectrum in Fig.~\ref{fig:nanozeeman}(a). Since there is no magnetic field, the fact that our measurements are being made on an ensemble of nanocrystals with different orientations has no real effect and we can proceed just as in the single crystal case. A best fit of the spectrum to a function of the form $I\left( \nu  \right)$ gives $C=0.04$, $E=0.0050\;\rm{GHz}$, $D=2.687\;\rm{GHz}$, and $\Gamma=0.012\;\rm{GHz}$.

\begin{figure}[!htbp]
\ifeps
  \includegraphics[width=3.25in]{nanodiamonds2.eps}
\else
  \includegraphics[width=3.25in]{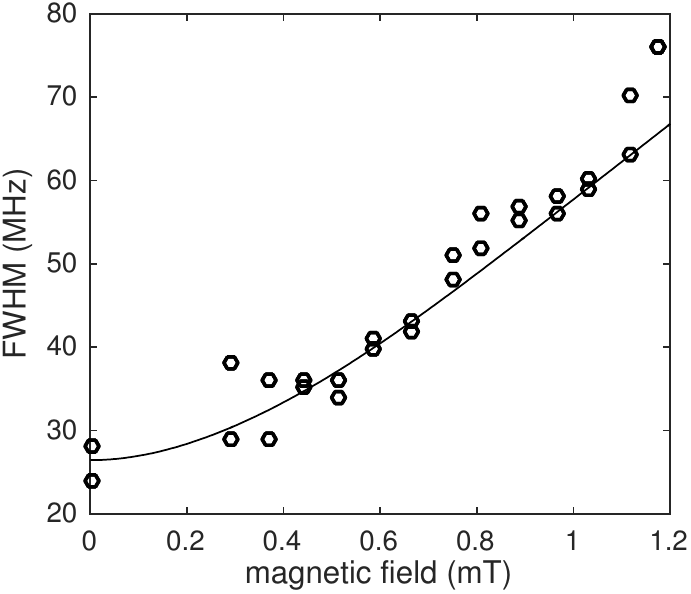}
\fi  
\caption{ Full width at half maximum of the ODMR resonance as a function of the magnetic field strength.The open circles represent values for the width derived from the  experimentally measured spectra while the black curve is generated from the model spectra.}
\label{fig:nanozeeman2}
\end{figure}

Figure~\ref{fig:nanozeeman}(b)--\ref{fig:nanozeeman}(d) shows ODMR spectra recorded in the presence of a static magnetic field $\mathbf{B}_0$, ranging in magnitude from 0.46 mT through 1.18 mT. The results are noticeably different from the single crystal case; we no longer observe distinct Zeeman levels as the field strength increases, but rather a steady broadening of the $\mathbf{B}_0 = 0$ resonance. This is because the crystal axes of the ensemble of nanocrystals are oriented in random directions with respect to $\mathbf{B}_0$. The gray curves are produced by extending the model used in the single crystal case by summing over all the different (assumed equally likely) orientations of the nanocrystals with respect to $\mathbf{B}_0$. There is qualitative agreement between the model and the measured spectra, but the model does not  account for the fact that the experimental spectra are not symmetrical about the center frequency $D$. The absence of well-defined troughs makes it harder to extract accurate values for  the magnetic field, but the width of the resonance can serve as an indicator of the strength of the applied magnetic field. Figure~\ref{fig:nanozeeman2} shows the full width at half maximum of the resonance as a function of the magnetic field strength. The open circles represent values for the width derived from the measured spectra while the black curve represent widths obtained from the model spectra.  This gives a simple means to translate the measured width of a resonance curve into a magnetic field strength, this time using ODMR spectra recorded from small quantities of nanocrystals. Having a simple method to optically measure magnetic field strengths with a spatial resolution of several microns has significance for biological applications, since diamond nanocrystals can be inserted into living cells.\cite{kucsko2013nanometre}

\subsection{ODMR in an instructional setting}
The Wellesley Physics Department's ``junior lab'' course is organized in a fairly standard format: We have a suite of advanced experiments that students rotate through.  Working in pairs over the course of about one and one-half weeks (three 3-hour sessions) students can complete a typical experiment.  Each experiment comes with a write-up that summarizes the theoretical background, suggests in general terms  an appropriate  experimental procedure, and then gives some guidance as to how to analyze the data. We try to give enough guidance so that our students can be successful, but not so much that it feels like they are simply following a recipe. The ODMR experiments described here fit well within this framework. Starting from a blank breadboard they construct the fluorescence microscope and then record ODMR spectra from a large single crystal. We then ask them to analyze the data in the framework of the  models presented above. Switching to a microcrystal sample presents additional challenges. Exciting and collecting  fluorescence from a single microcrystal requires a more careful optical alignment. Also, since the orientation of the microcrystal is not known in advance, the ODMR spectra recorded in the presence of an external magnetic field are typically more complicated and harder to interpret. (A challenging question for students is to see what they can infer about the orientation of the microcrystal from their measurements.) Overall the experience is a good mix of experimental technique, careful data analysis, and modeling based on physics that connects to what they have seen in their quantum mechanics class. A copy of the lab write-up for this experiment is included with the supplementary material.


In the Advanced Placement Physics course at Dover Sherborn High School, students cycle through a sequence of four modern experiments, each lasting two hours.  For the diamond magnetometer experiment, students use the simple set up described above to record ODMR spectra from an NV-rich large single crystal diamond. Using a strong permanent magnet to generate a magnetic field, they observe how the spectrum changes as they vary (by hand)  the distance from the magnet to the diamond. They make a video recording of this process, with a field of view that includes the spectrum and the permanent magnet, as well as a ruler taped to the table. By analyzing this video they can determine how the magnet's field varies with distance.

\section{Conclusion}
What makes a lab experience for advanced undergraduate physics majors both compelling and educational? Physicists have a long and valuable tradition of building their own instruments and many of the most important advances in scientific history were based on a combination of science, engineering, and design. But this tradition may be waning. Both the power and the problem with much modern scientific instrumentation is reflected in the term ``black box'' that is commonly used to describe the equipment. Today's black-box instruments are highly effective in making measurements and collecting data, enabling even novices to perform advanced scientific experiments. But, at the same time, these black boxes are ``opaque'' -- in that their inner workings are often hidden and thus poorly understood by their users. In contrast, the fluorescence microscope used in this experiment can be largely set up from scratch by students. There is some irony in the fact that, in order to eliminate interference from room light, the instrument that students construct is literally located inside a black box. But unlike a metaphorical black box, this is one they can reach inside, to build and to explore.
\vspace{20 pt}

\section{Appendix: Parts list}

Here, for convenience, we give a parts list for the experiments described above. We do not include in these lists commonly available parts such as oscilloscopes, function generators, and dc power supplies.

\subsection{Samples}

Sources for the samples used in these experiments are listed in Table ~\ref{samples}. For large NV-rich single crystals, suitable starting material is 2 point (that is 4 milligram) high-pressure high-temperature synthetic diamond single crystal from Element Six. It must then be electron irradiated and annealed, as described above.

\begin{table}[h!]
\centering
\caption{Samples}
\label{samples}
\begin{tabular}{|c|c|c|c|}
\hline
\textbf{Part}			& \textbf{Supplier}		& \textbf{Part}			& \textbf{Cost(\$)}	\\ \hline
microcrystals			& Ad\'{a}mas			& MDNV15umHi50mg 	& 300			\\ \hline
nanocrystals			& Ad\'{a}mas			& ND-NV-100nm		& 300			\\ \hline
singlecystals			& Element Six			& Monocrystal 2pt		& 125-400			\\ \hline
\end{tabular}
\end{table}

\subsection{Microscope parts}
The optical components and mounts for the fluorescent microscope are listed in Table ~\ref{microscopeparts}. This is generally research  grade equipment, but there are a number of lower cost alternatives that can work reasonably well. For example, the \textit{xyz} stage listed below is nice, but it is expensive and the precision it provides is greater than what is required here.  An adequate  replacement at 1/4 the cost can be obtained from banggood.com, part number 1105874. The same goes for the microscope objective: just about any 10x objective will do. The photodiode with adjustable gain amplifier could easily be replaced by a 50 cent raw photodiode and a home made current to voltage converter.

\begin{table}[h!]
\centering
\caption{Microscope parts}
\label{microscopeparts}
\begin{tabular}{|c|c|c|c|}
\hline
\textbf{Part}			& \textbf{Supplier}	& \textbf{Part}		& \textbf{Cost(\$)}	\\ \hline
10$times$ microscope objective	& Thorlabs		& RMS10X		& 340			\\ \hline
thread adapter			& Thorlabs		& SM1A3			& 13				\\ \hline
dichroic mirror 			& Thorlabs		& DMLP550 		& 170			\\ \hline
long pass filter			& Thorlabs 		& FEL0600		& 75				\\ \hline
short pass filter			& Thorlabs 		& FES0750		& 75				\\ \hline
ND filter wheel			& Thorlabs 		& FW1AND		& 300			\\ \hline
camera			& Thorlabs 		& DCC1645C		& 355			\\ \hline
100 mm camera lens		& Thorlabs 		& MVL100M23		& 185			\\ \hline
photodiode detector		& Thorlabs 		& PDA36A		& 320			\\ \hline
mirror				& Thorlabs 		& ME1-G01		& 13				\\ \hline
1'' optics holders (3)		& Thorlabs 		& LRM1			& @15			\\ \hline
1" pedestal posts (5)		& Thorlabs 		& RS1P8E		& @22			\\ \hline
0.5" pedestal post		& Thorlabs 		& RS0.5P8E		& 22				\\ \hline
clamping forks (6)		& Thorlabs 		& CF125C			& @11			\\ \hline
flip mount				& Thorlabs 		& TRF90			& 82				\\ \hline
\textit{xyz} stage		& Newport 		& MS-125-XYZ		& 580			\\ \hline
construction rails (4)		& Thorlabs		& XE25L09		& @15 			\\ \hline

\end{tabular}
\end{table}

\subsection{Laser parts}
The laser system we used was expensive (\$2600) but the Coherent Compass 215M is just as good at 1/5 the price. (Downsides: The Compass 215M is intended as an OEM system and is not available directly from the manufacturer. It is however readily available online -- eBay, for example.  Also, you will need to provide a 5V, 4A power supply.) The power stability these systems provide (1\%) improves the signal to noise ratio, but it is not absolutely essential provided you do fast scans. The Laserglow system is stable to only about 10\% but it is even less expensive and is plug and play.

The lasers used in these experiments are Class IIIB and eye safety must be taken very seriously. There are a number things we do to make things as safe as possible for novice student users. The first line of defense is the requirement that everyone wear appropriate laser safety glasses (optical density at least 2.0 at 532 nm) while the laser is on. We also place  an absorptive neutral density filter with an optical density of at least 2.0  immediately in front of the laser during the alignment phase of the experiment, when laser beams are most likely to go astray. Finally, the black box enclosure for the entire system keeps the beams confined to a small area in the room.   

\begin{table}[h!]
\centering
\caption{Laser Parts}
\label{laserparts}
\begin{tabular}{|c|c|c|c|}
\hline
\textbf{Part}			& \textbf{Supplier}		& \textbf{Part}		& \textbf{Cost(\$)}	\\ \hline
laser safety glasses		& DiOptika			& LG-005L		& 50				\\ \hline
40 mW laser module 	& Thorlabs 			& DJ532- 40 		& 180			\\ \hline
laser mount 			& Thorlabs 			& LDM21 			& 340			\\ \hline
current source			& Thorlabs			& LDC210C 		& 1100			\\ \hline
temperature  controller	& Thorlabs			& TED200C 		& 1000			\\ \hhline{|=|=|=|=|}
50 mW laser system       	& Coherent 			& Compass 215M 	& 500			\\ \hhline{|=|=|=|=|}
20 mW laser system		& Laserglow			& LCS- 532 		& 300			\\ \hline
\end{tabular}
\end{table}

\subsection{Microwave parts}

External dc power supplies are required for both the voltage controlled oscillator (5V) and the microwave amplifier (12V). 

\begin{table}[h!]
\centering
\caption{Microwave parts}
\label{microwaveparts}
\begin{tabular}{|c|c|c|c|}
\hline
\textbf{Part}				& \textbf{Supplier}	& \textbf{Part}		& \textbf{Cost(\$)}	\\ \hline
voltage controlled osc.		& Mini-Circuits		& ZX95-3150+		& 40				\\ \hline
microwave amplifier 			& Mini-Circuits		& ZRL-3500 		& 140			\\ \hline
2 dB attenuators (4)			& Omni Spectra 	& 2082-6171-02	& @15			\\ \hline
\end{tabular}
\end{table}

\section{Acknowledgments}
This work relied heavily on the support of our colleagues at the \textit{Center for Integrated Quantum Materials}. In particular, we wish to thank Nathalie de Leon for first suggesting this project and for many subsequent discussions that were essential to our progress. The development of the simplified setup greatly benefited from help from Michael Walsh and Hannah Clevenson as well as Marko Lon\v{c}ar, Anna Schneidman, Robert Hart, John Free, and Danielle Braje. Daniel Twitchen and Mathew Markham at Element 6 graciously provided the large single crystal samples. Jim MacArthur from the the Harvard Physics Electronics Shop built the high resolution microwave source, based on a design by Sasha Zibrov, while Paul Horowitz made some useful suggestions on the design of the low resolution microwave source. Wellesley students Catherine Matulis, Hanae Yasakawa, Phyllis Ju and Hannah Peltz Smalley were instrumental in getting our diamond studies at Wellesley off the ground. This work is supported by the Center for Integrated Quantum Materials under NSF grant DMR-1231319. We are also grateful for support from Wellesley College, including the Sally Etherton Cummins Summer Science Research Endowed Fund and the T.T. and W.F. Chao Summer Scholars Program in Natural Sciences Endowed Fund. 

\ifbibtex
\bibliography{ODMR} 

\begin{thebibliography}{99}
\bibitem{footenote1} This is the specification for the Bruker 500 MHz NMR spectrometer that is the workhouse used in organic chemistry labs at Wellesley College.
\bibitem{footnote3} A number of factors are responsible for the weakness of the interaction of the spin with its surroundings. For example, because the $^{12}\rm{C}$ nucleus has a spin of zero, the ``spin-spin'' interactions are small. Also, the low mass and stiff interatomic bonds of the carbon lattice lead to a high Einstein temperature ($T_E = 2300 \rm{K}$), which makes the interaction of the NV center) with the vibrational modes of the surrounding lattice unusually weak at room temperature.
\bibitem{footnote4} Equation \ref{eq:gyro} is only valid in the limit that the magnetic field strength is much smaller than the zero-field splitting between the $m_s=0$ state and the $m_s=\pm1$ states; i.e when $B_{||} << 2.87 \rm{GHz}/\gamma$. This condition is satisfied for all the experiments described here.
\bibitem{footnote6} Excellent lower cost alternative laser systems are the Coherent Compass 215M or the Laserglow Technolgies LCS-532 series. There are even less expensive 532 nm lasers, but these are typically not temperature stabilized. This results in significant fluctuations in output power that may obscure the relatively small variations in fluorescence intensity that we observe here. To reduce the detrimental effects of power fluctuations it helps to record the spectra quickly. Also, for the laser intensities used here, the fluorescence is linearly proportional to the laser power and for longer scans we have had good success in minimizing the effect of laser power fluctuations by monitoring the laser power and normalizing.
\end{thebibliography}

\else

\fi

\end{document}